\documentclass[11pt,a4paper]{article}

\usepackage[margin=1in]{geometry}
\usepackage{amsmath,amssymb}
\usepackage{booktabs}
\usepackage{hyperref}
\usepackage{xcolor}
\usepackage{tikz}
\usetikzlibrary{quantikz}

\hypersetup{
  colorlinks=true,
  linkcolor=blue,
  citecolor=blue,
  urlcolor=blue
}

\providecommand{\ket}[1]{\left|#1\right\rangle}
\providecommand{\bra}[1]{\left\langle#1\right|}

\newcommand{\Ry}{R_y}
\newcommand{\CNOT}{\mathrm{CNOT}}

\title{\textbf{Closed-Form Optimal Quantum Circuits for Single-Query Identification of Boolean Functions}}
\author{Leonardo Bohac}
\date{\today}

\begin{document}
\maketitle

\begin{abstract}
We study minimum-error identification of an unknown single-bit Boolean function given black-box (oracle) access with one allowed query.
Rather than stopping at an abstract optimal measurement, we give a fully constructive solution: an explicit state preparation and an
explicit measurement unitary whose computational-basis readout achieves the Helstrom-optimal success probability $3/4$ for
distinguishing the four possible functions. The resulting circuit is low depth, uses a fixed gate set, and (in this smallest setting)
requires no entanglement in the input state.

Beyond the specific example, the main message is operational. It highlights a regime in which optimal oracle discrimination is not only
well-defined but \emph{implementably explicit}: the optimal POVM collapses to a compact gate-level primitive that can be compiled,
verified, and composed inside larger routines. Motivated by this, we discuss a ``what if'' question that is open in spirit: for fixed
$(n,m,k)$, could optimal $k$-query identification (possibly for large hypothesis classes) admit deterministic, closed-form descriptions
of the inter-query unitaries and the final measurement unitary acting on the natural $n{+}m$-qubit input--output registers (and, if
needed, small work registers)? Even when such descriptions are not compact and do not evade known circuit-complexity barriers for
generic Boolean functions, making the optimum constructive at the circuit level would be valuable for theory-to-hardware translation and
for clarifying which forms of ``oracle access'' are physically meaningful.
\end{abstract}

\section{Motivation and contribution}

In the oracle model of quantum algorithms, an oracle call is typically treated as a unit-cost primitive.
That abstraction is powerful, but it leaves open a practical question that matters whenever one wants to
\emph{build} an oracle-discrimination routine rather than merely analyze it: what is the optimal strategy,
and can it be written down as a small, explicit circuit?

We focus on the smallest complete-identification task: the four Boolean functions on one input bit.
Unlike Deutsch's promise problem (constant vs.\ balanced) \cite{deutsch_jozsa}, our goal is to \emph{identify which function it is}
after a single query, allowing a nonzero probability of error.

This task can be framed as \emph{unitary/oracle discrimination}: a single application of an unknown oracle maps a chosen probe state to one of several candidate output states, and the goal is to infer which oracle was applied via an optimal measurement. This perspective has been used to analyze oracle-assisted algorithms more broadly (for example, the Deutsch--Jozsa setting as discrimination among unitary transformations) \cite{collins_dj}. Related work also studies distinguishing \emph{sets} of Boolean functions via state-filtering/generalizations of Deutsch--Jozsa \cite{bergou_herzog_hillery_prl,bergou_hillery}, and broader questions of oracle-operator discrimination and its limitations \cite{chefles_oracle}. Our focus is complementary: minimum-error identification of the complete four-element one-bit oracle family with a single query, together with a closed-form, gate-level realization of an optimal measurement. The central message is that in this setting the
optimum is not only attainable but \emph{constructive}: it admits closed-form expressions for both the optimal
probe preparation and the optimal measurement unitary, together with a compact gate decomposition.

This is more than a convenient presentation. Many minimum-error discrimination problems are solvable ``in principle''
(the optimal POVM exists and can be characterized, e.g., by convex optimization), yet the resulting measurement is
not naturally expressed as a reusable gate-level primitive. \emph{Our result suggests a different regime exists}: (i) the abstract optimum is fixed by symmetry and can be written down, and (ii) the corresponding measurement can be compiled into a small, explicit circuit. The optimality mechanism (square-root/pretty-good measurements for symmetric or geometrically uniform ensembles) is well understood; the contribution here is to pin down an explicit closed form for this oracle family and to provide a fully constructive realization. In particular,
a regime in which the optimum is fixed by symmetry and therefore yields a concrete circuit that can be compiled,
verified, and composed as a subroutine inside larger algorithms.

Concretely, we provide:
\begin{itemize}
  \item the optimal one-query success probability, $P_\star = 3/4$;
  \item an optimal probe state $\ket{\phi}$ that is \emph{separable} (no entangling gate is needed for preparation);
  \item a closed-form two-qubit unitary $U_1$ that realizes the optimal measurement as a projective computational-basis
        measurement, together with a standard decomposition using two CNOTs and four $\Ry$ rotations.
\end{itemize}

Knowing such a closed-form, compact gate composition is \emph{not cosmetic --- it is a systems-level advantage}:
it turns ``there exists an optimal measurement'' into an auditable circuit with predictable cost, enabling direct
resource estimates (depth and two-qubit gate count), hardware scheduling, and straightforward inclusion in
benchmarking and verification workflows.

\section{Setting: four one-bit Boolean oracles}

Let $f:\{0,1\}\to\{0,1\}$. There are four possibilities:
\[
f_0(x)=0,\quad
f_1(x)=1,\quad
f_2(x)=x,\quad
f_3(x)=1-x.
\]
We are given black-box access to $f$ through the standard quantum oracle
\begin{equation}
O_f \ket{x}\ket{y} = \ket{x}\ket{y\oplus f(x)}, \qquad x,y\in\{0,1\}.
\label{eq:oracle}
\end{equation}
We assume the oracle is chosen uniformly from $\{f_0,f_1,f_2,f_3\}$ and may be queried once.

A one-query discrimination protocol has the form
\[
\ket{00}\xrightarrow{U_0}\ket{\phi}\xrightarrow{O_f}\ket{\psi_f}:=O_f\ket{\phi}\xrightarrow{U_1}
\text{measure in the computational basis},
\]
followed by the decision rule ``if the measured basis label is $m\in\{0,1,2,3\}$, output $f_m$ as the guess.''

For clarity, we use the standard computational-basis ordering
\(\ket{0}=\ket{00},\ket{1}=\ket{01},\ket{2}=\ket{10},\ket{3}=\ket{11}\),
so the decision rule is equivalently the decoding table below:
\begin{center}
\begin{tabular}{c c c}
\hline
Measured state & Label \(m\) & Reported function \\ \hline
\(\ket{00}\) & 0 & \(f_0\) \\
\(\ket{01}\) & 1 & \(f_1\) \\
\(\ket{10}\) & 2 & \(f_2\) \\
\(\ket{11}\) & 3 & \(f_3\) \\ \hline
\end{tabular}
\end{center}

With this convention, the average success probability is
\begin{equation}
P_{\mathrm{succ}}(U_0,U_1)
=\frac{1}{4}\sum_{m=0}^{3}\left|\bra{m}\,U_1\,O_{f_m}\,\ket{\phi}\right|^2,
\qquad \ket{\phi}=U_0\ket{00}.
\label{eq:psucc}
\end{equation}
Our goal is to maximize $P_{\mathrm{succ}}$ over all two-qubit unitaries $U_0,U_1$.

\section{Why the optimum is computable in closed form}

\subsection{A symmetry that fixes the optimal measurement}

Let $G=\{I,\ I\otimes X,\ \CNOT_{01},\ (I\otimes X)\CNOT_{01}\}$ acting on two qubits.
For a fixed probe state $\ket{\phi}$, the four post-oracle states can be written as
\[
\ket{\psi_g} = U_g\ket{\phi},\qquad U_g\in G,
\]
which makes the ensemble \emph{geometrically uniform} (generated from a single state by a finite unitary group).
For equiprobable geometrically uniform pure-state ensembles, the square-root measurement (also called the
least-squares measurement or ``pretty-good measurement'') minimizes the probability of error and can be realized
as a projective measurement after a suitable unitary preprocessing \cite{barnett_croke_review,hausladen_wootters,eldar_forney_srm,krovi_symmetric}.
This reduces the optimization to finding a probe state whose induced Gram matrix maximizes the square-root
measurement success probability.

\subsection{The optimal success probability}

For our ensemble, the optimal Gram matrix can be chosen as
\begin{equation}
G^\star=
\renewcommand{\arraystretch}{1.2}
\begin{pmatrix}
\phantom{-}1 & -\tfrac{1}{3} & \phantom{-}\tfrac{1}{3} & \phantom{-}\tfrac{1}{3}\\
-\tfrac{1}{3} & \phantom{-}1 & \phantom{-}\tfrac{1}{3} & \phantom{-}\tfrac{1}{3}\\
\phantom{-}\tfrac{1}{3} & \phantom{-}\tfrac{1}{3} & \phantom{-}1 & -\tfrac{1}{3}\\
\phantom{-}\tfrac{1}{3} & \phantom{-}\tfrac{1}{3} & -\tfrac{1}{3} & \phantom{-}1
\end{pmatrix},
\label{eq:gram_opt}
\end{equation}
whose eigenvalues are $(4/3,4/3,4/3,0)$.
For an equiprobable geometrically uniform ensemble, the square-root measurement success probability can be expressed
in terms of the Gram matrix eigenvalues (equivalently, via the trace of $\sqrt{G}$) \cite{barnett_croke_review,hausladen_wootters,eldar_forney_srm,krovi_symmetric,helstrom}.
Plugging the eigenvalues of \eqref{eq:gram_opt} yields
\[
P_\star = \frac{1}{16}\left( 3\sqrt{\frac{4}{3}} \right)^2 = \frac{3}{4}.
\]
In the next section we give an explicit probe $\ket{\phi}$ and a concrete circuit that achieves $P_\star$.

\section{An explicit optimal circuit}

\subsection{A separable optimal probe state}

An optimal probe can be chosen as
\begin{equation}
\ket{\phi}
= a\ket{00}+b\ket{01}+a\ket{10}+b\ket{11},
\qquad
a=\frac{1+\sqrt{2}}{2\sqrt{3}},
\quad
b=\frac{1-\sqrt{2}}{2\sqrt{3}}.
\label{eq:phi_opt}
\end{equation}
The equalities in \eqref{eq:phi_opt} imply a product structure:
\[
\ket{\phi}
=\frac{1}{\sqrt{2}}(\ket{0}+\ket{1})
\;\otimes\;
\big(\sqrt{2}a\,\ket{0}+\sqrt{2}b\,\ket{1}\big),
\]
so state preparation needs no entangling gate. One convenient choice is
\begin{equation}
U_0 = H\otimes \Ry(\theta_0),
\qquad
\theta_0=2\arcsin\!\left(\frac{\sqrt{2}-2}{2\sqrt{3}}\right),
\label{eq:U0}
\end{equation}
which satisfies $U_0\ket{00}=\ket{\phi}$.

\subsection{A closed-form measurement unitary}

Define
\[
\gamma=\frac{1}{2\sqrt{2}},
\qquad
\alpha=\frac{1}{2}+\gamma,
\qquad
\beta=\frac{1}{2}-\gamma.
\]
An optimal preprocessing unitary (followed by computational-basis measurement and the rule ``outcome $m$ means $f_m$'')
is
\begin{equation}
U_1=
\begin{pmatrix}
\gamma & -\gamma & \alpha & \beta \\
-\gamma & \gamma & \beta & \alpha \\
\alpha & \beta & -\gamma & \gamma \\
\beta & \alpha & \gamma & -\gamma
\end{pmatrix}.
\label{eq:U1_matrix}
\end{equation}
With $\ket{\psi_m}=O_{f_m}\ket{\phi}$, this choice achieves
\[
\frac{1}{4}\sum_{m=0}^3\left|\bra{m}U_1\ket{\psi_m}\right|^2=\frac{3}{4}.
\]
One standard decomposition uses two CNOTs and single-qubit rotations:
\begin{equation}
U_1=
(I\otimes H)\,
(\Ry(\theta_3)\otimes \Ry(\theta_4))\,
\CNOT_{01}\,
(\Ry(\theta_1)\otimes \Ry(\theta_2))\,
\CNOT_{10}\,
(I\otimes H),
\label{eq:U1_decomp}
\end{equation}
with
\[
\theta_1=-\frac{3\pi}{4},\quad
\theta_2=-\frac{\pi}{2},\quad
\theta_3=\frac{\pi}{4},\quad
\theta_4=\frac{\pi}{2}.
\]

\subsection{End-to-end circuit and gate count}

The full protocol is shown below. The only two-qubit gates outside the oracle are the two CNOTs in $U_1$.

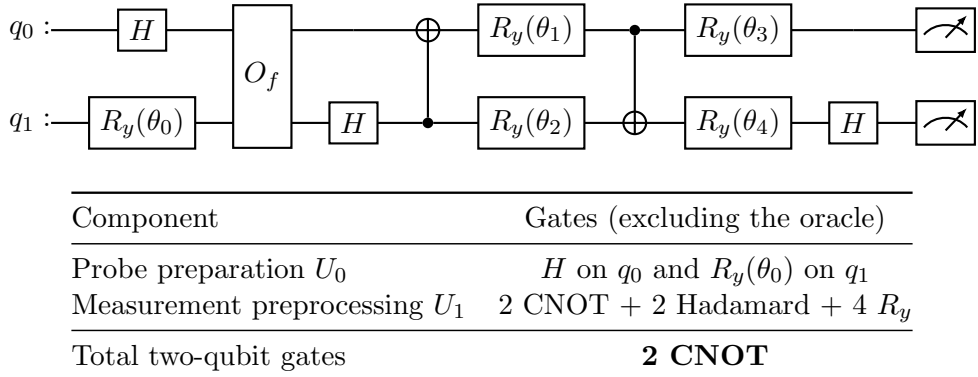
\begin{center}
\begin{quantikz}
q_0: & \gate{H} & \gate[2]{O_f} & \qw & \targ{} & \gate{\Ry(\theta_1)} & \ctrl{1} & \gate{\Ry(\theta_3)} & \qw & \meter{} \\
q_1: & \gate{\Ry(\theta_0)} & & \gate{H} & \ctrl{-1} & \gate{\Ry(\theta_2)} & \targ{} & \gate{\Ry(\theta_4)} & \gate{H} & \meter{}
\end{quantikz}
\end{center}

\begin{center}
\begin{tabular}{@{}lc@{}}
\toprule
Component & Gates (excluding the oracle) \\
\midrule
Probe preparation $U_0$ & $H$ on $q_0$ and $\Ry(\theta_0)$ on $q_1$ \\
Measurement preprocessing $U_1$ & 2 CNOT + 2 Hadamard + 4 $\Ry$ \\
\midrule
Total two-qubit gates & \textbf{2 CNOT} \\
\bottomrule
\end{tabular}
\end{center}

\section{Implications and outlook}
\subsection{What changes --- and what remains open}

In the oracle model of quantum algorithms, a query is treated as a unit-cost primitive. In hardware, however, an oracle must be
synthesized as a circuit, and for large input sizes almost all Boolean functions have no succinct implementation
\cite{shannon_switching}. This tension is often glossed over in purely query-based analyses, yet it becomes central when the goal is to
\emph{build} an oracle-discrimination routine rather than merely reason about it.

Our contribution is to make the optimal strategy fully explicit in the smallest nontrivial instance: the entire minimum-error protocol
reduces to two closed-form unitaries (preparation and measurement), with a transparent two-qubit gate count and depth. This is a
\emph{concrete conceptual advance} because it separates two notions that are often conflated:
(i) the existence of an optimal measurement as an abstract POVM, and
(ii) the existence of an optimal \emph{circuit} that realizes it in a form suitable for compilation and deployment.
In this setting the optimal POVM can be taken to be a projective measurement after a fixed preprocessing unitary, so optimality becomes
an executable circuit object rather than an implicit optimization output.

\subsection{Why closed-form gate compositions are more than cosmetic}

Closed-form gate compositions enable capabilities that a purely POVM-level statement does not:
\begin{itemize}
  \item \textbf{Composability.} A unitary-plus-basis-readout measurement can be embedded as a drop-in subroutine in larger routines
        (e.g., discrimination inside a learning loop, verification protocol, or amplitude-amplification wrapper) without re-solving an
        optimization problem at each use.
  \item \textbf{Compilability and accounting.} A compact circuit has a transparent hardware cost (two-qubit gate count, depth, and
        rotation parameters), enabling fault-tolerant resource estimates and device-level scheduling analysis. This turns an information
        measure into an implementable, schedulable primitive.
  \item \textbf{Verification and benchmarking.} ``Optimal'' becomes experimentally testable: the circuit provides a minimal, verifiable
        instance of optimal minimum-error oracle discrimination that can be executed and stress-tested on real devices.
\end{itemize}

A notable aspect of the present construction is that it uses only the natural $n{+}m$-qubit input--output registers for the oracle (here
two qubits) and does not require additional workspace. Whether such ``minimal-register'' realizations persist in broader identification
settings is itself a meaningful design question.

\subsection{A ``what if'' question: closed-form strategies beyond the smallest case}

The example in this paper suggests a broader possibility worth taking seriously, while keeping feet on the ground: for fixed integers
$(n,m,k)$, one can always \emph{define} the optimal $k$-query identification problem for the hypothesis class of Boolean functions
$f:\{0,1\}^n\!\to\!\{0,1\}^m$, and the optimal success probability is, in principle, attainable by some choice of inter-query unitaries and
a final measurement \cite{helstrom,barnett_croke_review}. What is usually missing is an explicit circuit-level description.

\emph{What if there were simple formulas?} Concretely: could there exist deterministic, closed-form prescriptions for the sequence
$U_0,\ldots,U_k$ and a final unitary realizing the optimal measurement (via Naimark dilation), as a function of $(n,m,k)$? Such a result
would be impactful even if the resulting circuits are not compact for large $n$. It would (a) turn an optimization-defined object into a
uniform, reproducible circuit primitive, (b) enable systematic resource accounting for query-optimal identification, and (c) potentially
reveal unexpected representation-theoretic or symmetry structure in families that, a priori, look unstructured. In other words, it would
further extend the ``constructive regime'' perspective: optimality would come with a deterministic recipe rather than only existence.

At the same time, known circuit-complexity barriers remain relevant. In particular, implementing a \emph{given} oracle for a generic
Boolean function requires exponential resources \cite{shannon_switching}, so closed-form prescriptions would not, by themselves, turn
arbitrary oracles into efficiently realizable hardware primitives. The more realistic expectation is that explicit formulas, if they
exist, will either (i) yield circuits whose size reflects this exponential cost, or (ii) apply cleanly to broad but structured subclasses
where both oracle synthesis and discrimination have exploitable symmetry (e.g., group-covariant families, geometrically uniform state
ensembles \cite{eldar_forney_srm,krovi_symmetric}).

\subsection{Outlook}

From a practical standpoint, the next steps are concrete: extend the fully constructive analysis to the next smallest regimes (e.g.,
$(n,m,k)=(2,1,1)$ or $(1,1,2)$), and use these cases to test whether compact, minimal-register measurement unitaries persist or whether
ancillas become necessary. From a theoretical standpoint, it is natural to pursue two complementary directions: (i) derive explicit
closed-form solutions for larger \emph{structured} families (where symmetry is expected to fix the optimal measurement), and (ii) identify
and formalize obstructions that explain when such explicit circuit descriptions cannot remain compact.
\section{Conclusion}

Complete identification of a Boolean oracle is a natural strengthening of Deutsch's one-bit classification task.
For the four one-bit Boolean functions, the minimum-error one-query optimum admits a fully explicit solution:
$P_\star=3/4$, a separable optimal probe state, and a compact two-CNOT measurement circuit.

Beyond its pedagogical clarity, this example highlights a broader point. Many oracle-discrimination optima can be stated
only implicitly (``an optimal POVM exists''), while here symmetry yields a constructive optimum: a closed-form circuit
that can be compiled, verified, and reused. In this sense, the result points to a structured regime of multi-query
identification problems in which analytic characterization and compact implementations are possible, and it clarifies
that such closed forms are signatures of structure rather than a route to synthesizing arbitrary (random) Boolean
oracles.

\section*{Acknowledgments}
The author acknowledges the broader quantum information community for foundational contributions to optimal quantum state discrimination and square-root measurement techniques.

\end{document}